\documentclass[12pt]{article}

\author{Vladimir A. Petrov \footnote{e-mail:Vladimir.Petrov@ihep.ru}}

\title{On Sizes of Hadrons\footnote{The abridged version of the talk at the XXXVII Int'l Workshop on HEP, July 22–24, 2025, Protvino, Moscow region, Russia.} }
\date{}

\begin{document}

\maketitle
A. A. Logunov Institute for High Energy Physics, NRC KI,
Protvino, RF

\begin{abstract}

\section*{Abstract}

 Recently, various types of "radii" have been intensively discussed in the literature: "charge", "mass", "mechanical", "gravitational", etc.
In my report, I analyze the definitions of quantities of such type in terms of matrix elements of local field operators and their relationship to physical (geometric) size. I also try to find out possible (if any) interpretation of the mentioned  quantities and point out some serious conceptual difficulties related to the causality principle.
\end{abstract}

\section*{Introduction}

The lion's share of results in particle physics are functions 
of Lorentz-invariant combinations of four-momenta. 

However, spatio-temporal characteristics are also present.
For example, resonance lifetimes. 

In eikonal schemes for amplitudes, models are formulated in terms of impact parameters (transverse coordinates). 

Among the usual set of particle characteristics in Particle Data Group publications, such as mass, spin, and the like, so-called "charge radii" appear. Below, we will analyze the physical meaning of these quantities and establish their connection with local hadron operators and the causality principle.

The "mean square charge radius" of the proton, $ r_p $ is defined as (see, e.g.\cite{mil})
\begin{equation}
r^2_p=6dG_E/dq^2\mid_{q=0}
\end{equation}
where $ G_E(q^2) $ stands for the Sachs electric form factor. 
Concerning the mathematical meaning of $ r_p^2 $, this is not a normal "average" as it is thought to be an integral of some "charge density" which doesn't define a probabilistic (positively defined) measure.
It cannot also be considered as a measure of the spatial extent of a charge particle because the slope $ dG_E/dq^2\mid_{q=0} $ is not always positive (e.g. for the neutron).
The reason of the sign indefiniteness is certainly the presence of charges of constituents having different signs.
Simple quark model considerations allow \cite{Ptr} to retrieve the geometric size of the proton (if we ignore the subtleties that are discussed in the following sections)in terms of the valence quark spatial distributions. The physical average size of the proton
turns out to be different from  the PDG value
\[r^2_p= (0.84... fm)^2\]
and, in the isotopic invariance approximation, reads
\[r^2_ {p, phys} = (0.77...fm)^2.\]
The physical size of the neutron turns out, in this approximation, to be equal to that of the proton, while from the
slope of the form factor we would have
\[r^{2}_n= - 0.116 fm^2.\]

Note that we are ignoring massless particles in this context. Although massless hadrons have not yet been observed, this case is nevertheless relevant to our topic, as will be discussed in more detail below.

\section*{Radius definition in the QFT context.}

All radii mentioned above are defined as a slope of some from-factor related to one or another current operator ( electromagnetic current,baryon current, energy-momentum tensor etc.).
In this Section we will use a generic form factor $G(q^2) $
without specification of its nature because our interest will be to inquire a space-time content of the standard radius definition
\begin{equation}
r^2:=6dG(q^2)/dq^2\mid_{q=0}
\end{equation}
By its name, the quantities of the type $ r^{2} $ refer to three-dimensional space, but according to Eq. (2), it is also a Lorentz invariant. Since we are talking about the "radius," an isotropic quantity, we can reconcile these two aspects only by recognizing that we are talking about the radius of a hadron in the rest frame.
To see if this is the case let us use the Bogoliubov reduction techniques\cite{Bog}  which introduces explicitly the space-time variables
which we are going to identify.
According to this procedure we get the following expression for the form factor
\begin{equation}
G(q^2) = \frac{1}{4m^{2}-q^2}\int d^{4}x \exp (-iqx) R(x,p)
\end{equation}
where $ R(x,p) $ is defined by the matrix elements of some 
bilocal causal operator. 

Coming back for concreteness to the electromagnetic  current $ J_\mu $ and a scalar target $ \vert p\rangle = a^+(p)\vert \Omega\rangle $ ($ \vert \Omega\rangle $ stands for the vacuum state)
\begin{equation}
R(x,p)= \langle\Omega\vert \delta J_\mu(x)/\delta\varphi^+(0)\vert p\rangle= i\theta (-x^0)\langle\Omega\vert [J_\mu(x),I^{+}(0)\vert p\rangle 2p^\mu
\end{equation}
where 
\begin{center}
$ I^{+}(y)= i(\delta S/\delta\varphi(y)) S^{+} = (\partial^{2}+ m^{2})\Phi(y)$ 
\end{center}
and $ \Phi(y) $ is the Heisenberg operator ($ \varphi $ is its asymptotic out-field ) corresponding to the target.
Bogoliubov microcausality condition reads:]
\begin{equation}
R(x,p)= 0, 0 > x^{0}> -\mid \textbf{x} \mid.
\end{equation}
i.e. outside the past light cone.
Now we have the following expression for the "radius"(squared)

\begin{equation}
r^2 = \int d^{4}x \Re R(x,p)[\frac{(xp)^2}{m^2} - x^2] 
\end{equation}

Two observation are worth to make. 
First, the quantity $ r^2 $ is not an average because $ R(x,p) $ is not positive definite although it is real.
Second,the "averaged" expression $ [...] $ in Eq.(6) only in the target's rest frame is the spatial radius.
\begin{center}
$  [(xp)^2/m^{2} - x^2]\mid_{\textbf{p}=0}= \textbf x^{2}. $
\end{center}

So, one could think of a completely innocent-looking expression

\begin{equation}
r^{2} = \int d^{4}x\textbf{x}^{2} \rho(\textbf{x}) )
\end{equation}
where 
\[\rho (\textbf{x}) = \int dx^{0}\Re R(x,p =(m,\textbf{0})).\]
Moreover, as it should be, $ \rho (\textbf{x}) $ has the property of a bona fide charge density:

\[\int d\textbf{x}\rho (\textbf{x})= Q\]
where $ Q $ is the charge of the state $\vert p\rangle  $
\[\hat{Q}\vert p\rangle = \int d\textbf{x}J_{0}(x)\vert p\rangle=Q\vert p\rangle.\]
Recall that the "charge density"$ \rho (\textbf{x}) $ is not and should not be sign-definite.

So, everything seems to look fine: we obtained an expression for the charge spatial distribution in terms of local hadron operators.
This, in turn, could provide a solid basis for concrete calculations and approximations.

Unfortunately, not everything is so rosy.

\section*{What is the generically defined hadron radius from a relativistic perspective?}

In view of the well-known relativity of lengths and time, we will consider the problem from a clearly relativistic point of view.
Recall that in the relativity theory \cite{Log} the proper length of a body is defined as follows.For simplicity we will take a one-dimensional body.
The interval squared between the infinitesimally close 4-coordinates of the ends A,B of the body is ($ x^{\mu}= x^{\mu}_{A}-x^{\mu}_{B} $)

\begin{equation}
ds^{2} = g_{\mu\nu}dx^{\mu}dx^{\nu}= g_{00}(dx^{0})^{2}+2g_{0j}dx^{0}dx^{j}+g_{ij}dx^{i}dx^{j}
\end{equation}

The spatio-temporal quantities in Eq.(8) change when the reference frame changes. However, the invariant interval $ ds^{2}$ can be expressed in terms of physical, frame independent quantities, i.e.
\begin{equation}
ds^{2} = c^{2}d\tau^{2}- dr^{2}
\end{equation}
where
\[cd\tau = g_{0i}dx^{i}/\sqrt{g_{00}}\]
and
\[dr^{2} = [(g_{0i}g_{0k}/g_{00}) - g_{ik}] dx^{i}dx^{k}.\]

\[i,k = 1,2,3.\]
The proper length (proper size) of an object is the distance between its ends, measured in the reference frame in which the object is at rest. Physically, this is realized by measuring the coordinates of the ends simultaneously (at equal times) in this inertial reference frame.
In our terms we get for the proper length of the object
\[dr^{2} = - ds^{2} \] at \[d\tau =0.\]

It can be proven that in the integrand of Eq.(6) the time difference $ t=x^{0}/c $ vanishes only in the non-relativistic limit
\begin{equation}
R(x^{0}=ct,\textbf{x},p)\mid_{c\rightarrow\infty}\sim \delta (t)f(\textbf{x}), 
\end{equation}
and we get an instant snapshot of the object.
In the general case, even in the laboratory frame, we obtain a quantity "smeared over time." Moreover, the interval $ x^{2} $ in the integrand of Eq.(7) lies, due to the causality principle, within the (backward) light cone and cannot be made space like in any reference frame.An attempt to place $ x^{\mu} $ on a space like surface ($ x^{0} = 0$,in particular)leads to vanishing of our $ \rho (\textbf{x}) $.
\section*{Conclusion}
In this talk I have attempted to highlight some important issues that arise when analyzing the existing definition of charge (and other) radii. 

In general, this is connected with the long-standing problem of localization in relativistic quantum field \cite{Heg} theory and apparently requires some new concepts that reflect the undoubted finiteness of the concentration region of any physical  characteristics of a particle.

It should be noted that a critical analysis of a  possibility of consistently defining quantities such as three-dimensional "charge density" can also be found in paper \cite{Mil}, although its author addresses the issue from a completely different perspective.


\begin{thebibliography}{99}

\bibitem {mil}

Gerald A. Miller,

\textit{Defining the proton radius: A unified treatment.}

Phys.Rev.C 99 (2019) 3, 035202, 

e-Print: 1812.02714 [nucl-th].

\bibitem{Ptr}

V.A. Petrov and  V.A. Okorokov,

\textit{The size seems to matter or where lies the "asymptopia"?}

Int.J.Mod.Phys.A 33 (2018) 13, 1850077, 

e-Print: 1802.01559 [hep-ph].

\bibitem{Bog}
N.N. Bogolyubov, B.V. Medvedev, M.C. Polivanov,

\textit{Problems of the theory of dispersion relations}

(1956) Princeton, NJ by Inst. Adv. Stud.; 

N.N.Bogoliubov and D.V. Shirkov,

\textit{Introduction to the theory of quantized fields,}

1959 New York : Interscience Publishers.

\bibitem {Log}

Anatoly Logunov,

\textit{Lectures In Relativity And Gravitation – A Modern Look.}

Nauka Publishers, Moscow ,1991.

\bibitem{Heg}

Gerhard C. Hegerfeldt,

\textit{Remark on Causality and Particle Localization.}

Phys.Rev.D 10 (1974) 3320.

\bibitem {Mil}
Gerald A. Miller,

\textit{Impossibility of obtaining time-independent, three-dimensional, spherically symmetric densities of confined systems of relativistically moving constituents.}

Phys.Rev.C 112 (2025) 4, 045204,
e-Print: 2507.14388 [hep-ph].


\end{thebibliography}
\end{document}